# Computing the mass of a magnetic monopole in SU(2) gauge theory[*]

J. Smit[a] and A.J. van der Sijs[b]

[a] Institute for Theoretical Physics, University of Amsterdam,
Valckenierstraat 65, 1018 XE Amsterdam, The Netherlands
[e-mail: `jsmit@phys.uva.nl`]

[b] Department of Theoretical Physics, University of Zaragoza,
Facultad de Ciencias, 50009 Zaragoza, Spain
[e-mail: `arjan@sol.unizar.es`]

A single magnetic monopole in pure SU(2) gauge theory is simulated on the lattice and its mass is computed in the full quantum theory. The results are relevant for our proposed realization of the dual superconductor hypothesis of confinement.

## 1. INTRODUCTION

In Ref. [1] we presented a realization of the dual superconductor hypothesis of confinement. The monopoles were of the 't Hooft-Polyakov (HP) form, in particular the Bogomol'nyi-Prasad-Sommerfield (BPS) solution. Assuming that the configurations considered dominate the string tension $\sigma$ we obtained the estimate $\sqrt{\sigma}/\Lambda_{\overline{MS}} = 2.3$.

In our model we made two assumptions. The first one was that quantum fluctuations screen the (chromo-)electric field of the classical BPS monopole, turning it into what we call an HP-like monopole. The second assumption concerned the mass of the monopole. The mass of the classical BPS monopole is $M = 4\pi\mu/g^2$, where $\mu$ is the scale characterizing the solution. We conjectured that the mass of the monopole in the full quantum theory behaves as

$$M = \frac{4\pi\mu}{g_R^2(\Lambda_R/\mu)} C(g_R^2(\Lambda_R/\mu)) \ . \qquad (1)$$

Here $g_R$ is the running coupling in the $R$-scheme defined in terms of the quark-antiquark potential, with $\Lambda_R/\Lambda_{\overline{MS}} = 1.048$ [2]. The function $C$ was supposed to be slowly increasing as for the HP monopole, $1 < C \lesssim 2$.

The subject of the present study, described in detail in Ref. [3] (see Ref. [4] for preliminary work), is the assumption about the mass of the monopole. We put a single monopole background in a box and use lattice Monte Carlo methods to include quantum fluctuations. The monopole background is created by imposing appropriate monopole boundary conditions. We use the simulation data to determine the function $C(g_R^2(\Lambda_R/\mu))$ in order to discuss the validity of the mass formula (1).

## 2. THE MAGNETIC MONOPOLE

The classical magnetic monopole in euclidean pure SU(2) gauge theory is given by

$$A_k^a(\vec{x},t) = \epsilon_{akl}\hat{x}_l \frac{1 - K(\mu r)}{r} \ , \qquad (2)$$

$$A_4^a(\vec{x},t) = \delta_{ak}\hat{x}_k \frac{H(\mu r)}{r} \ , \qquad (3)$$

with

$$H(\mu r) = \mu r \frac{\cosh \mu r}{\sinh \mu r} - 1 \ , \qquad (4)$$

$$K(\mu r) = \frac{\mu r}{\sinh \mu r} \ , \qquad (5)$$

as in the BPS limit of the HP monopole. The scale parameter $\mu$ is arbitrary and can be regarded as an inverse core size of the monopole. The configuration (2–5) is a static solution to the equations of motion, with mass $M = 4\pi\mu/g^2$.

---
[*]Presented by A.J. van der Sijs

We want to study the monopole in a 4-volume consisting of a finite, non-periodic spatial box of size $(2R)^3$ times a periodic time direction of extent $T$. The idea is to induce the monopole by fixing the fields in the boundary at a value suggested by the asymptotic behaviour of the monopole field. We will assume $\mu R > 1$, so that effects exponential in $\mu R$ are suppressed.

The appropriate boundary conditions for the BPS monopole are given by its asymptotic behaviour

$$H(\mu r)/r \sim \mu - \frac{1}{r(\vec{x})}, \qquad (6)$$

$$K(\mu r) \sim 0. \qquad (7)$$

However, if our assumption is correct that the monopole becomes HP-like due to quantum fluctuations, 'HP-like' asymptotic behaviour, $H/r \sim \mu$, will be more appropriate at semi-classical values of the coupling.

We take HP-like boundary conditions with parameter $\mu_0$ for our dynamical simulations, i.e. $K \sim 0$, $H/r \sim \mu_0$. Although BPS-like and HP-like boundary conditions are not equivalent because $1/r(\vec{x})$ varies along the boundary of the cubic box, an analysis of classical monopole energies shows that our boundary conditions are compatible with both an HP monopole of scale $\mu = \mu_0$ and a BPS monopole of scale $\mu_{\text{eff}} = \mu_0 + 1/R_{\text{eff}}$.

Another important result of the classical analysis is that there is a symmetry implying that $\mu_0$ boundary conditions are equivalent with monopoles characterized by scales $\mu_0' = \mu_0 + 2\pi n/T$, for arbitrary integer $n$. This symmetry also allows monopoles of opposite electric charge, characterized by $\mu_0 < 0$, to show up among the monopoles at positive values of $\mu_0$. All these monopole configurations are local minima of the action, but only the global minimum is important in the simulations. As a consequence, the accessible range of $\mu_0$ values is restricted to $0 < \mu_0 \leq \pi/T$. We choose $\mu_0 = \pi/T$ to minimize the influence of nearby local minima.

## 3. THE MONOPOLE MASS

The monopole mass can be written as

$$M = M_{\text{in}} + M_{\text{out}}, \qquad (8)$$

with

$$M_{\text{in}} = -\frac{1}{T} \ln \frac{Z_{\text{mon}}(\beta; a\mu_0)}{Z_{\text{vac}}(\beta)} \qquad (9)$$

the contribution from inside the box, measured in the simulation, and $M_{\text{out}}$ a correction term for the outside region. Here

$$Z_{\text{mon}}(\beta; a\mu_0) = \int DU \exp[-S_{\text{plaq}}(U; \beta)] \qquad (10)$$

is the partition function subject to monopole boundary conditions and $Z_{\text{vac}}(\beta)$ is the analogous definition for vacuum boundary conditions, i.e. $A_\mu^a = 0$.

Eq. (9) can be written in a form accessible to Monte Carlo computation by differentiating it with respect to $\beta$ and subsequently integrating again,

$$M_{\text{in}}(\beta) = \int_0^\beta \frac{d\tilde{\beta}}{\tilde{\beta}} \Delta E(\tilde{\beta}). \qquad (11)$$

Here $\Delta E$ is the 'internal energy'

$$\Delta E(\beta) = \frac{1}{T}(\langle S \rangle_{\text{mon}} - \langle S \rangle_{\text{vac}}). \qquad (12)$$

In order to compute $M_{\text{in}}$ using Eq. (11) the integral is replaced by a sum. At each value of $\tilde{\beta}$ in this summation two simulations are needed, to compute $\langle S \rangle$ with both monopole and vacuum boundary conditions. High statistics is required to compute the difference of the two large numbers in Eq. (12) accurately.

Fig. 1 shows the internal energy for a simulation at an $8^4$ lattice, with $a\mu_0 = \pi/8 = 0.39$. In the low $\beta$ region, $\Delta E \approx 0$ because the interior of the box is decorrelated from the boundary. At intermediate $\beta$ a monopole is induced in the box, and for large $\beta$ the classical energies are approached. The interesting region extends from $\beta = 2.3$ onwards, where one expects scaling behaviour and the physical value of $\mu$ increases from the order of the string tension scale $\sqrt{\sigma}$ to the high-momentum region.

The $\Delta E$ data are integrated to obtain $aM_{\text{in}}(\beta)$. Subsequently, $C_{\text{in}}(g_R^2(\Lambda_R/\mu))$ is found as follows. First $\Lambda_R/\mu$ is calculated using Monte Carlo data for $a\sqrt{\sigma}$ [5] and an input value for $\sqrt{\sigma}/\Lambda_R$ (we used three different values [1,6,7]).

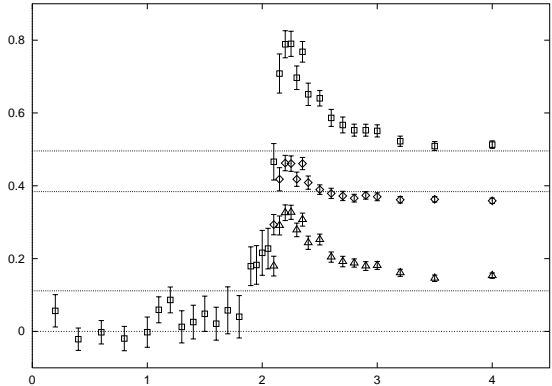

Figure 1. $\Delta E$ (in units of $4\pi/ag^2$) as a function of $\beta = 4/g^2$. Shown are the magnetic ($\diamond$) and electric ($\triangle$) components and the total ($\square$). The horizontal lines denote the classical values.

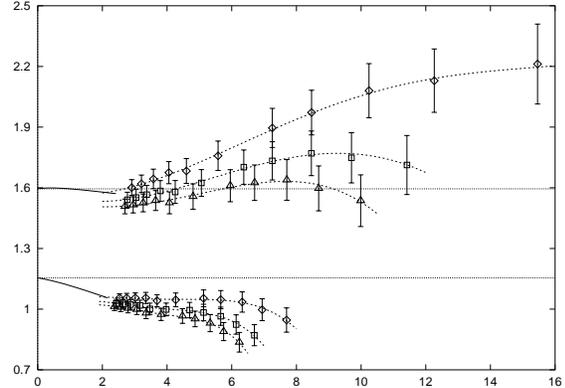

Figure 2. $C$ as a function of $g_R^2(\Lambda_R/\mu)$. The different sets of points are for $\sqrt{\sigma}/\Lambda_R = 1.7$ ($\diamond$), 2.0 ($\square$) and 2.2 ($\triangle$), for both $\mu = \mu_0$ (upper set of curves) and $\mu = \mu_{\text{eff}}$ (lower set). The horizontal lines denote the classical limit ($g^2 = 0$). The solid curves come from large-$\beta$ fits of the mass data.

Next $g_R^2(\Lambda_R/\mu)$ is calculated using its two-loop $\beta$-function and the corresponding $C_{\text{in}}$ is extracted from $aM_{\text{in}}$ using Eq. (1). This procedure is carried out for both $\mu = \mu_0$ and $\mu = \mu_{\text{eff}}$, corresponding to HP-like and BPS-like behaviour of the monopole, respectively.

The result for $C$, including the correction for the exterior region, is shown in Fig. 2. For $g_R^2 \approx 0$ we know the monopole is BPS-like, so the HP analysis (upper three curves) is misleading there. At larger couplings, the monopole may or may not become HP-like. If it does, the $C$ values will lie in the region indicated by the upper set of curves. This means that $C$ increases from $C = 1$ at weak coupling to $C \approx 1.6$ at $g_R^2 \approx 8$ or $C \approx 2.0$ at $g_R^2 \approx 10$, depending on the value of $\sqrt{\sigma}/\Lambda_R$. This is in good agreement with our assumptions. If, however, the monopole remains BPS-like at large coupling, our first assumption does not apply. Nevertheless, even in that scenario (lower set of curves) $C$ remains almost constant, $C \approx 1$ up to $g_R^2 \approx 6$.

**Acknowledgements:** This work was carried out when AvdS was at the University of Oxford, supported by SERC (U.K.) grant GR/H01243. He is currently sponsored by DGICYT (Spain). JS is supported by the Stichting FOM. The numerical simulations were performed on the Cray Y-MP4/464 at SARA with support from the Stichting NCF.